**This is the accepted manuscript (postprint) of the following article:**
S. Shahroudi, A. Parvinnasab, E. Salahinejad, S. Abdi, S. Rajabi, L. Tayebi, *Efficacy of 3D-printed chitosan-cerium oxide dressings coated with vancomycin-loaded alginate for chronic wounds management*, Carbohydrate Polymers, 349 (2025) 123036.
https://doi.org/10.1016/j.carbpol.2024.123036

# Efficacy of 3D-Printed chitosan-cerium oxide dressings coated with vancomycin-loaded alginate for chronic wounds management

Sharareh Shahroudi [a,1], Amir Parvinnasab [a,1], Erfan Salahinejad [a,2], Shaghayegh Abdi [b], Sarah Rajabi [b,c], Lobat Tayebi [d,e]

[a] *Faculty of Materials Science and Engineering, K. N. Toosi University of Technology, Tehran, Iran*

[b] *Department of Tissue Engineering, School of Advanced Technologies in Medicine, Royan Institute, Tehran, Iran.*

[c] *Department of Cell Engineering, Cell Science Research Center, Royan Institute for Stem Cell Biology and Technology, ACECR, Tehran, Iran*

[d] *Marquette University School of Dentistry, Milwaukee, WI 53233, USA*

[e] *Institute for Engineering in Medicine, Health, & Human Performance (EnMed), Batten College of Engineering and Technology, Old Dominion University, Norfolk, VA 23529, USA*

## Abstract

Multifunctional wound dressings with antibacterial and antioxidant properties hold significant promise for treating chronic wounds; however, achieving a balance of these characteristics while maintaining biocompatibility is challenging. To enhance this balance, this study focuses on the design and development of 3D-printed chitosan-matrix composite scaffolds, which are incorporated with varying amounts of cerium oxide nanoparticles (0, 1, 3, 5, and 7 wt%) and subsequently coated with a vancomycin-loaded alginate layer. The structure, antibiotic drug delivery kinetics, biodegradation, swelling, biocompatibility, antibacterial, antioxidant, and cell migration behaviors of the fabricated dressings were evaluated in-vitro. The findings reveal that all of the formulations demonstrated a robust antibacterial effect against *S. aureus* bacterial strains in disk diffusion tests. Furthermore, the dressings containing cerium oxide nanoparticles exhibited proper antioxidant capabilities, with over 78.1% reactive oxygen species (ROS) scavenging efficiency achieved with 7% cerium oxide nanoparticles. The sample containing 5% cerium oxide nanoparticles was identified as the optimal formulation, characterized by the most favorable cell biocompatibility, an ROS scavenging ability of over 73.4%, and the potential to close the wound bed within 24 h. This study highlights that these dressings are promising for managing chronic wounds by preventing infection and oxidative stress in a correct therapeutic sequence.

[1] These authors contributed equally and share co-first authorship.

[2] Corresponding author: Email Address: <salahinejad@kntu.ac.ir>

## 1. Introduction

Chronic wounds are a type of injury that fails to heal within the expected timeframe, frequently remaining unhealed for extended periods ranging from weeks to months or even years. Conventional wound dressings are insufficient for treating chronic wounds due to the lack of some principal requirements [1,2]. Alternatively, multifunctional wound dressings loaded with various therapeutic agents, including antimicrobials, growth factors, antioxidants, and anti-inflammatory agents are more effective for this purpose. They reduce the risk of infection, minimize the need for frequent dressing changes, regulate moisture levels, promote oxygenation, and reduce the need for multiple types of dressings [3,4].

Antioxidant and antibacterial properties are essential for the treatment of chronic wounds [5,6]. Zeng *et al.* [7] introduced epigallocatechin-3-gallate and copper-functionalized silk fibroin hydrogel dressings with antioxidant and antibacterial properties. Hassan *et al.* [8] explored antibacterial and antioxidant chitosan-hyaluronan-phosphatidylcholine dihydroquercetin wound dressings. However, these dressings did not demonstrate a significant balance of biocompatibility, antibacterial, and antioxidant properties, highlighting the need for improved design. It is known that the release of antibacterial agents should be rapid and burst within initial hours, followed by the sustained release of antioxidants to ensure an optimal healing process for chronic wounds [6,9]. To the best of our knowledge, wound dressings containing antioxidant and antibacterial agents with such controlled and sequential release have not been developed for chronic wounds up to date. Accordingly, in this paper, we introduce a new design to address such dressings, relying on three-dimensional (3D) printing as an ideal technique to produce multifunctional, customized, patient-specific wound dressings. These dressings are characterized by chitosan scaffolds incorporated with cerium oxide nanoparticles and then coated with a vancomycin-loaded alginate layer.

Chitosan is a semi-synthetic biopolymer derived from naturally occurring chitin that is found in the exoskeletons of crustaceans such as crabs and shrimp. Chitosan has various desirable properties, such as biocompatibility, biodegradability, and antimicrobial activity, making it ideal for wound dressings [10,11]. Chitosan wound dressings effectively promote wound healing by creating a moist environment, which facilitates tissue regeneration. Chitosan dressing also acts as a barrier against infection by absorbing excess wound exudate and preventing bacterial growth. They also have a high absorption capacity, which can reduce the changing frequency of dressings. Several studies have demonstrated the effectiveness of chitosan 3D-printed wound dressings in promoting wound healing [12–15]. However, these dressings may not be sufficient for the effective treatment of chronic wounds, which require considerable antibacterial and antioxidant properties.

Several studies have investigated the application of cerium oxide ($CeO_2$) nanoparticles in wound healing due to their unique properties, particularly the ability to act as antioxidants and anti-inflammatory agents [16,17]. Oxidative stress can harm the tissue and impede the healing process. Cerium oxide nanoparticles have the potential to scavenge oxidative stress, which in turn can aid wound healing by decreasing inflammation and encouraging the development of new tissues [18]. Studies show that hydrogel dressings containing cerium oxide nanoparticles enhance wound healing through their anti-inflammatory and antioxidant activities [19–21], but the risk of infection needs additional consideration.

Antibiotic-loaded wound dressings can benefit wounds at high risk of infection, such as burns, surgical, or chronic wounds. By releasing antibiotics directly into the wound bed, these dressings can help prevent the growth of bacteria and reduce the risk of infection [22]. Numerous studies have been carried out on the efficacy of wound dressings that contain vancomycin, both for the prevention of wound infections and for the treatment of wounds that are already infected [23–25]. Vancomycin is commonly prescribed for serious infections caused by gram-positive bacteria, including *Staphylococcus aureus* (*S. aureus*) and *Enterococcus faecalis*. This antibiotic works by inhibiting bacterial cell wall synthesis, which prevents the bacteria from growing and reproducing. However, its release kinetics should be accurately controlled using appropriate local delivery platforms.

Alginate is another polysaccharide biopolymer derived from brown seaweed with wide use in wound dressings. Alginate can absorb large amounts of water and wound exudate, making it an effective wound care material. When alginate comes into contact with wound exudate, it forms a hydrophilic gel that conforms to the shape of the wound, providing a moist environment to promote wound healing [26]. Alginate has been utilized in diverse forms for wound dressing applications, including coating on chitosan. El-feky *et al.* [27] reported that alginate-coated chitosan nanogels loaded with silver sulfadiazine demonstrated significant therapeutic efficacy for the treatment of burn wounds. Chen and colleagues [28] developed alginate-coated chitosan membranes for guided skin tissue regeneration applications by regulating soft tissue growth. These studies underscore the advantageous role of alginate coatings in chronic wound dressings, especially when combined with antimicrobial agents to tackle the issue of infection, leveraging their capability for effective burst drug delivery [4]. Typically, Aslani *et al.* [29] reported appropriate antibacterial properties of polylactide tissue engineering scaffolds after coating them with vancomycin-containing alginate. Similarly, Zhang *et al.* [30] developed vancomycin-containing alginate coatings on zeolite scaffolds with an initial burst release of vancomycin from the coatings, leading to potent antibacterial properties.

The hypothesis of this work is that 3D-printed chitosan scaffolds containing cerium oxide nanoparticles with a vancomycin-alginate coating offer an enhanced balance of biocompatibility, antibacterial, and antioxidant properties for chronic wounds management.

More specifically, the present research pioneers the development of multifunctional dressings with the sequential controlled release of antibacterial and antioxidant agents.

## 2. Materials and methods

### *2.1. Materials*

Chitosan (deacetylation degree: 84.2%, molecular weight: 140,000), sodium alginate (medium viscosity, molecular weight: 100,000, mannuronate/guluronate ratio: 1.56), vancomycin hydrochloride ($C_{66}H_{75}Cl_2N_9O_{24}.HCl$), cerium (III) chloride heptahydrate ($CeCl_3.7H_2O$), potassium hydroxide (KOH), phosphate-buffered saline (PBS, pH = 7.4), tetracycline ($C_{22}H_{24}N_2O_8$), 1-diphenyl-2-picrylhydrazyl ($C_{18}H_{12}N_5O_6$, DPPH), dimethyl sulfoxide ($C_2H_6OS$, DMSO), and formazan ($C_{18}H_{17}N_5S$), and 3-(4,5-dimethylthiazol-2-yl)-5-(3-carboxymethoxyphenyl)-2-(4-sulfophenyl)-2H-tetrazolium ($C_{20}H_{20}N_5O_6S_2$, MTS) were obtained from Sigma Aldrich. Glacial acetic acid ($CH_3COOH$), methanol ($CH_3OH$, 99%), and ammonia solution ($NH_4OH$, 25%) were also purchased from Merck.

### *2.2. Synthesis of cerium oxide*

The method used for preparing cerium oxide particles was based on a previously published paper [31]. Briefly, $CeCl_3.7H_2O$ was dissolved in 20 mL of distilled water under continuous magnetic stirring (Pole Ideal Pars, PIT 300) for 30 min to obtain a 0.67-M solution. 10 mL of the ammonia solution was then added dropwise while stirring until full precipitation. The precipitates were washed and collected by centrifugation (Pars Azma) at 4000 rpm for 10 min, repeated 10 times. Finally, the resulting powders were subjected to freeze-drying (Jalteb, JD 200) at -60 °C for 24 h.

### *2.3. Fabrication of chitosan-cerium oxide scaffolds*

For the preparation of 3D-printing inks, chitosan was dissolved in an acetic acid solution (2% v/v) under mechanical stirring (Teif Azma Teb, TAT-2500) for 2 h, resulting in a 9% w/v chitosan solution. The synthesized cerium oxide powder was added to the chitosan solution at different weight percentages of 1, 3, 5, and 7 wt% with respect to dry chitosan. The inks were then loaded into a 304 stainless steel cartridge of a 3D printer (Abtin Teb Fanavar, Abtin 2) attached to a nozzle with an internal diameter of 0.1 mm. Square-shaped scaffolds of 1×1×0.1 $cm^3$ were produced by depositing the inks layer-by-layer onto a plate set at 25 °C with a feed rate of 100 cc/min. The resulting scaffolds were treated with a 2 wt% potassium hydroxide solution for ionotropic gelation for 5 min and then soaked in pure ethanol for 30 min to remove any remaining residuals. Finally, the 3D-printed chitosan scaffolds were dried at 25 °C for one day in a vacuum oven (TOB, DZF-6050).

### *2.4. Coating of 3D-printed scaffolds with an alginate-vancomycin layer*

Sodium alginate was dissolved in distilled water under magnetic stirring (Pole Ideal Pars, PIT 300) to achieve an alginate solution with a concentration of 1.5 % w/v. Vancomycin was added to the solution at 0.7 % w/v, adapted from Ref. [30] showing a suitable balance of biocompatibility and antimicrobial properties. The scaffolds were then dipped in the prepared coating solution and then dried at room temperature for one day, giving the samples named in Table 1.

*Table 1. Samples designation*

| Sample name | Chitosan (w/v) | Cerium oxide (wt%) | Alginate (w/v) | Vancomycin (w/v) |
|---|---|---|---|---|
| Chi-Alg | 9 | - | 1.5 | - |
| Chi-1Ce-Alg-Van | 9 | 1 | 1.5 | 0.7 |
| Chi--3Ce-Alg-Van | 9 | 3 | 1.5 | 0.7 |
| Chi-5Ce-Alg-Van | 9 | 5 | 1.5 | 0.7 |
| Chi-7Ce-Alg-Van | 9 | 7 | 1.5 | 0.7 |

### *2.5. Characterization of cerium oxide powder*

The microstructure of the synthesized cerium oxide powder was explored by field emission scanning electron microscopy (FE-SEM, TESCAN, MIRA3) at an accelerating voltage of 15 kV. The phase structure of the powder was also identified by X-ray diffraction (XRD, GNR APD-2000, $CuK_\alpha$ radiation, 40 kV).

### *2.6. Characterization of alginate-coated chitosan 3D-printed scaffolds*

#### *2.6.1. Structural characterization*

The morphology and chemical composition of the scaffolds were analyzed using FE-SEM (TESCAN, MIRA3) coupled with energy dispersive X-ray spectroscopy (EDS) at an accelerating voltage of 15 kV. Fourier transformed infrared spectroscopy (FTIR, AVATAR, Thermo) was also used in the 4000-400 $cm^{-1}$ range for spectral analysis.

#### *2.6.2. Degradability and swelling properties*

The *in-vitro* degradation of the scaffolds was investigated through immersion in PBS at 37 °C. The scaffolds were removed at different time points, dried at 50 °C for 1 h, and weighed. The weight loss percentage was determined as follows:

$$\text{Weigth loss percentage} = \left(\frac{W_i - W_f}{W_i}\right) \times 100 \qquad (1)$$

where $W_i$ and $W_f$ represent the initial and final weights of the scaffolds at each time, respectively.

The swelling degree of the scaffolds was also determined by immersing them in PBS at 37 °C. After each time point, excess water was removed from the surface of the scaffolds, and the samples were weighed. The swelling percentage of the samples was calculated as follows:

$$\text{Swelling percentage} = \left(\frac{W_s - W_i}{W_i}\right) \times 100 \qquad (2)$$

where $W_i$ is the initial weight of the scaffolds and $W_s$ is the swollen weight of the scaffolds at any time points.

### *2.6.3. In-vitro drug release profile*

The *in-vitro* release of vancomycin from the alginate-coated scaffolds was investigated by soaking the scaffolds of 1×1×0.1 $cm^3$ in 20 mL PBS in an incubator set at 37 °C. The vancomycin concentration was measured at different times using UV-visible spectroscopy (Bloorazma, AURORA) at 281 nm, where the samples were placed in fresh PBS for subsequent time points.

### *2.6.4. Antibacterial activity*

The antibacterial efficacy of the dressings against the *S. aureus* and *E. coli* bacterial strains was assessed using the disk diffusion method. In the initial step, 50 µL of bacterial suspensions with a density of $10^8$ CFU/mL were inoculated onto agar plates. These plates were then incubated at 37 °C for 24 h. The 1×1×0.1 $cm^3$ scaffolds were sterilized using 15-min exposure to UV, carefully positioned onto the agar plates, and incubated at 37 °C for 24 h. Tetracycline antibiotic was used as the positive control, and paper disks inoculated with sterile saline were used as the negative control. Finally, microbial inhibition zones around each scaffold were determined precisely.

#### ***2.6.5. Antioxidant activity assay***

The ability of the dressings to scavenge DPPH free radicals was assessed to determine their antioxidant activity. Initially, the scaffolds were soaked in PBS for 5 h to remove the majority of the alginate-vancomycin coating and highlight the effect of cerium oxide, followed by drying at 60 °C for 1 h. A solution of 0.1 mM DPPH dissolved in methanol was prepared, and the scaffolds were immersed in 5 mL of this solution at a concentration of 25 mg/mL. The solutions containing the scaffolds were incubated in the dark at room temperature for 1 h, and the absorbance of the DPPH solution was recorded at 517 nm using UV-visible spectroscopy (Multiskan Go, Thermofisher).

#### ***2.6.6. Cell metabolic activity assay***

After sterilization using UV irradiation and 70% ethanol, the samples were moved into 24-well plates. $5\times10^4$ human dermal fibroblast (HDF) cells/cm$^2$ were seeded onto the scaffolds and cultured in a proliferation medium, with the medium changed every two days. The MTS assay was performed to assess the metabolic activity of the cells. On days 1, 3, and 5, the scaffolds were transferred to new wells, and the MTS solution was added. The plates were then incubated for 3 h in the dark at 37 °C. Subsequently, the media were transferred into a 96-well plate. Finally, the intensity of the media was measured utilizing an ELISA plate reader (Thermo Scientific Multiscan Spectrum) with optical density determination at 490 nm.

#### *2.6.7. Scratch assay*

The scratch assay was used to evaluate the ability of the developed dressings for facilitating wound closure, based on cell migration to fill a scratched area [51]. $2\times10^4$ HDF cells were placed in a 6-well cell culture plate and allowed to reach full confluence at 37 °C. A sterile pipet tip was used to create uniform scratches in each well. Light microscopy (Olympus, BX51) equipped with an Olympus DP72 digital camera was utilized to capture images of the cellular gaps at different time points (0, 6, 12, and 24 h). ImageJ 1.46r software was employed to measure the scratch area and calculate wound closure degrees, using Eq. 3.

$$\text{Wound closure degree} = \frac{S_0 - S_t}{S_0} \times 100 \qquad (3)$$

where $S_0$ and $S_t$ represent the initial scratch area and the area measured at each time point, respectively.

#### *2.6.8. Statistical analysis*

The measurements were conducted in triplicate, with data presented as mean ± standard deviation. Statistical analysis was performed using a Student t-test for comparisons between two groups or a one-way analysis of variance (ANOVA) for comparisons involving more than two groups, followed by a proper post hoc test. A p-value of $< 0.05$ was considered statistically significant.

## 3. Results and discussion

### *3.1. Structural characterization of cerium oxide powder*

The FE-SEM image of the synthesized cerium oxide powder (Fig. 1a) indicates agglomerates with sizes ranging from 60 to 120 nm, each consisting of nanoparticles within the size range of 15 to 25 nm with relatively smooth surfaces free of cracks or defects. This suggests that the nanoparticles possess a considerable level of structural and mechanical stability suitable for tissue engineering and drug delivery applications [32].

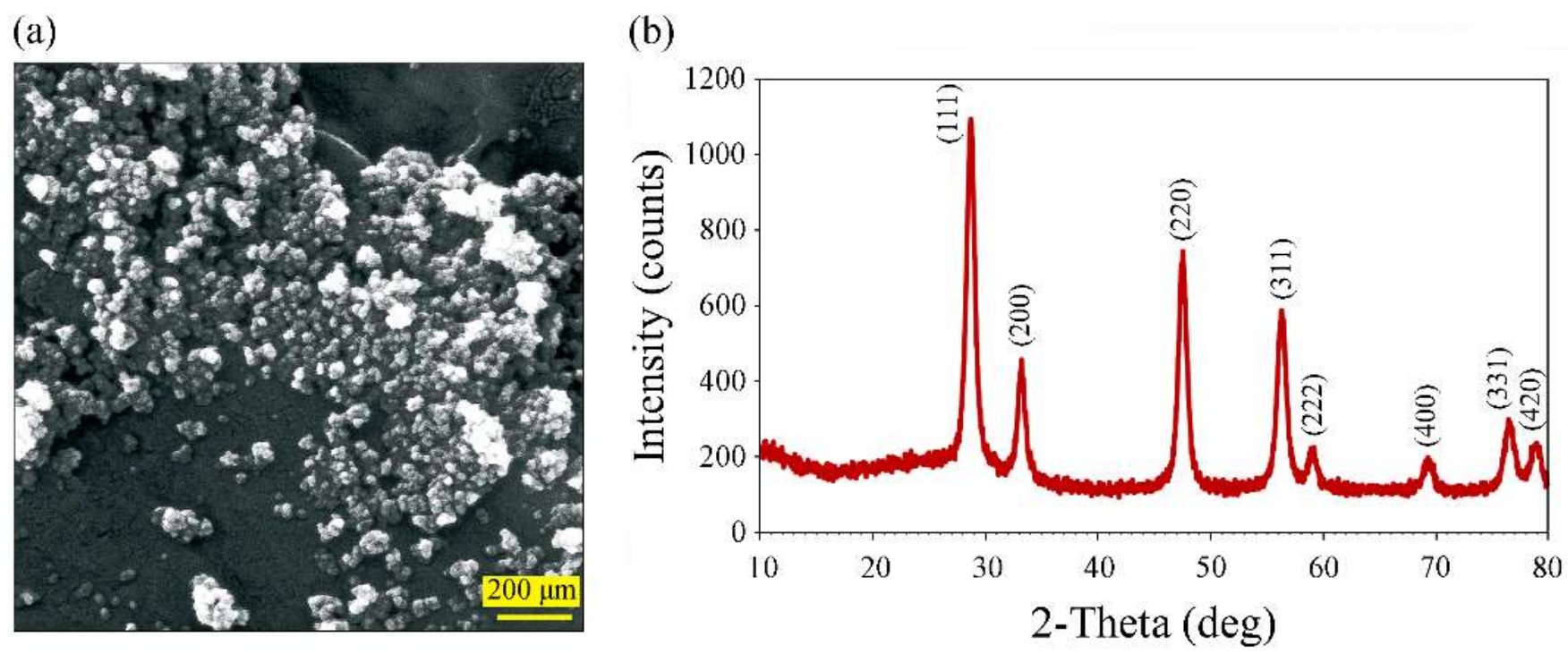

*Fig. 1. FE-SEM micrograph (a) and XRD patterns (b) of the synthesized cerium oxide powder.*

The XRD pattern of the synthesized cerium oxide nanoparticles is shown in Fig. 1b, presenting sharp and distinct peaks indicative of a high degree of crystallinity. The detected peaks correspond to the cubic fluorite structure of cerium oxide [33]. Furthermore, no secondary phase peaks are observed, indicating the high purity of the product. The broadening of the peaks is attributed to factors such as the small size of crystallites, lattice strain, and structural defects like vacancies and dislocations. The average crystallite size of 20 nm was estimated for the cerium oxide nanoparticles using the Scherrer equation, which is in good agreement with the FE-SEM observation. This nanometric size is promising for reactive oxygen species (ROS) scavenging ability because smaller particles contain more oxygen vacancies and $Ce^{3+}$ ions within their lattice, which are critical for antioxidant properties [34].

### *3.2. Structural characterization of dressings*

The FE-SEM images and EDS spectrum of a representative sample, Chi-5Ce-Alg-Van, are represented in Fig. 2. According to Fig. 2a, the 3D-printed scaffold has an interconnected porous structure with an average pore size of 700 microns. This porous feature is vital for cell activities as it allows for efficient oxygen and nutrient supply as well as waste removal [35]. Moreover, this structure aids in the removal of wound exudates, preventing bacterial accumulation and improving the wound healing process [36]. The extruded filaments in the constructs exhibit uniform sizes with an average diameter of 200 microns, making them suitable for conforming to wound bed sites, cellular infiltration, and tissue formation [37]. The uniformity of the alginate-vancomycin coating deposited on the 3D-printed filaments is observable in the cross-sectional image (Fig. 2b), with a mean thickness of approximately 3 microns. The other samples exhibited similar morphological features (micrographs are not provided) because all the samples contained a narrow range (0-7%) of the cerium oxide nanoparticles and were covered with the same alginate coating (except for Chi-Alg that was vancomycin-free). The relatively significant thickness and uniformity of the alginate coating across the samples contributed to the consistent topography observed. The EDS elemental analysis of Chi-5Ce-Alg-Van (Fig. 2c) shows the presence of carbon, oxygen, and nitrogen, which are the main constituting elements of alginate and vancomycin. Given the lower penetration depth of dispersive X-ray compared to the thickness of the alginate layer, the other samples showed the same EDS results.

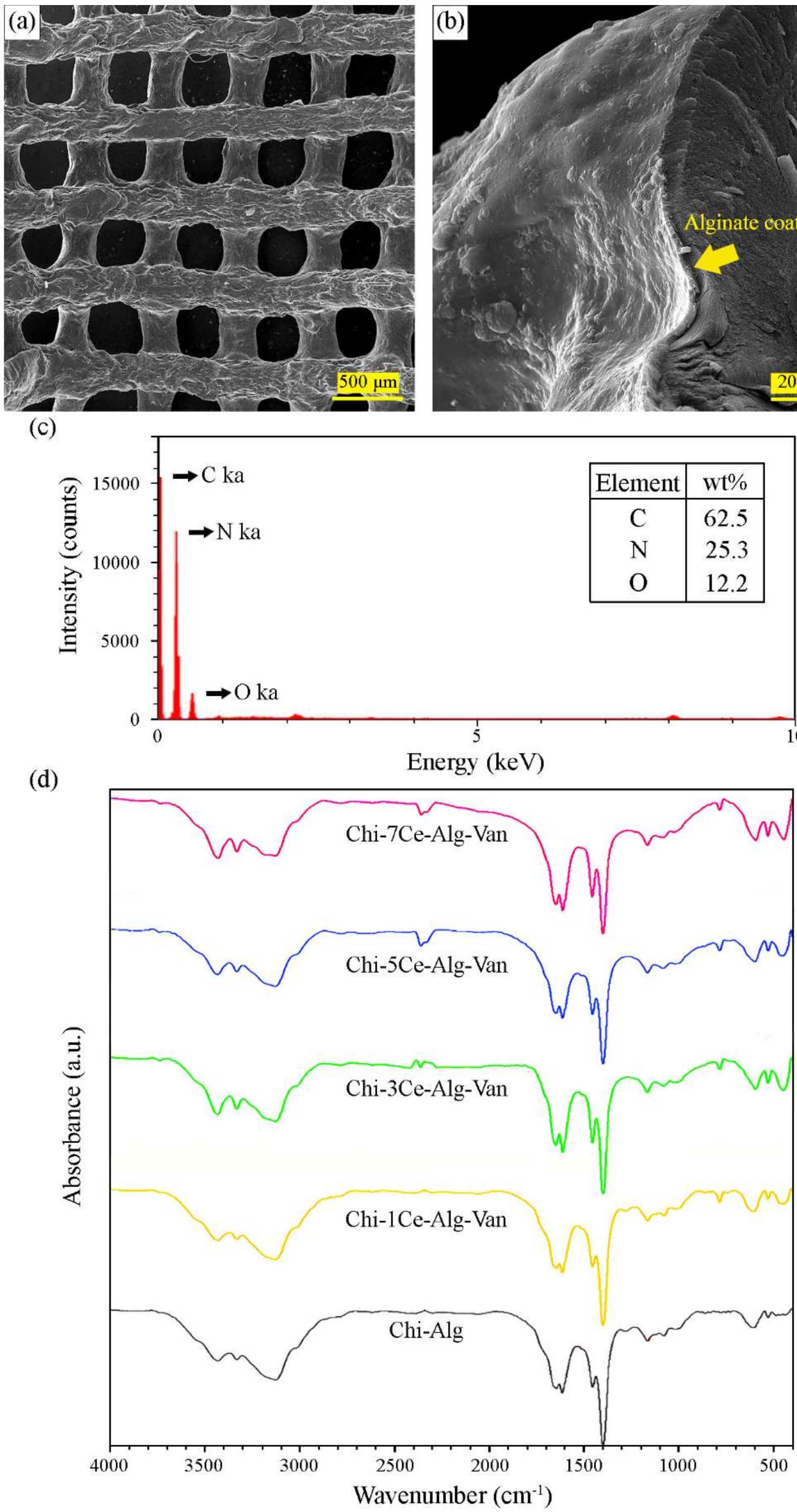
(a)
500 µm
(b)
Alginate coating
20 µm
(c)
C ka
N ka
O ka
Element wt%
C 62.5
N 25.3
O 12.2
Intensity (counts)
Energy (keV)
(d)
Chi-7Ce-Alg-Van
Chi-5Ce-Alg-Van
Chi-3Ce-Alg-Van
Chi-1Ce-Alg-Van
Chi-Alg
Absorbance (a.u.)
Wavenumber (cm-1)

*Fig. 2. Morphological FE-SEM micrograph (a), cross-sectional FE-SEM micrograph (b), and EDS spectrum (c) of the Chi-5Ce-Alg-Van sample. FTIR spectra of the different dressings (d).*

The FTIR spectra of the samples are indicated in Fig. 2d. Peaks located at 502-763 $cm^{-1}$ are attributed to Ce-O bonds, which correspond to the IR range (500-800 cm-1) of crystalline cerium oxide active phonon modes. The peak at 1466 $cm^{-1}$ is assigned to the rocking vibration of $-NH_2$, while peaks at 1114-1150 $cm^{-1}$ arise from the stretching vibration of C-N and C-O, endorsing the presence of chitosan and alginate, respectively. Absorbance peaks observed at 1263 and 1382 $cm^{-1}$ are attributed to the stretching vibration of chitosan main chains involving -C-O-C- and C-O bonds, respectively. The stretching vibrations of the chitosan OH bands are found at 3430, 1090, and 1030 $cm^{-1}$. Peaks observed at 1419 and 1466 $cm^{-1}$ are related to the stretching vibration of $CH_2$ bonds, while the peak at 1655 $cm^{-1}$ corresponds to the stretching vibration of C=O bonds. The broad band ranging from 3500 to 3100 $cm^{-1}$ closely match the stretching vibration of the N-H group bonded with the O-H group in chitosan, which aligns well with the literature [38]. As observed, by increasing the cerium oxide content, the bands associated with the stretching modes of OH and $NH_2$ become sharper, suggesting the interaction of chitosan with the cerium oxide nanoparticles within the $NH_2$ and OH groups [21]. The reduced intensity of the band located at 1419 $cm^{-1}$ in the presence of cerium oxide is indicative of an interaction between cerium oxide and the OH group linked to the $CH_2$ group in chitosan [39]. Regarding vancomycin, critical characteristic peaks include one at 1230 $cm^{-1}$ showing the presence of an aromatic ester, another at 1650 $cm^{-1}$ indicative of the skeletal vibration of the C-O bond, and a peak at 1062 $cm^{-1}$ representing the vibration of C-N [33,40]. Some of the mentioned peaks for vancomycin are somewhat obscured in the spectra due to overlaps with the characteristic peaks of chitosan and alginate in the 1200-1600 $cm^{-1}$ region. Nonetheless, the shifts observed in the peak positions and broadening suggest their presence. In conclusion, the FTIR analysis verifies the successful incorporation of all the intended constituents into the constructs.

### *3.3. Degradation and swelling behaviors of dressings*

The degradation and swelling behaviors of the dressings under the *in-vitro* physiological conditions are presented in Fig. 3. It is observed that adding cerium oxide up to 5% reduces the swelling and degradation rates of the samples, but the sample containing 7% cerium oxide exhibits higher rates. The influence of cerium oxide on the swelling and degradation of chitosan-based scaffolds has been previously studied, with two primary mechanisms being proposed. The first mechanism involves an increase in crystallinity due to the addition of cerium oxide, which results in lower swelling and degradation levels [41]. This mechanism appears to be predominant in the samples with the low amounts of cerium oxide. The second mechanism suggests that the release of cerium oxide from the scaffolds enhances porosity, favoring swelling and degradation [42]. This mechanism, with likely agglomeration at higher concentrations, explains the behavior observed in the sample containing 7% cerium oxide.

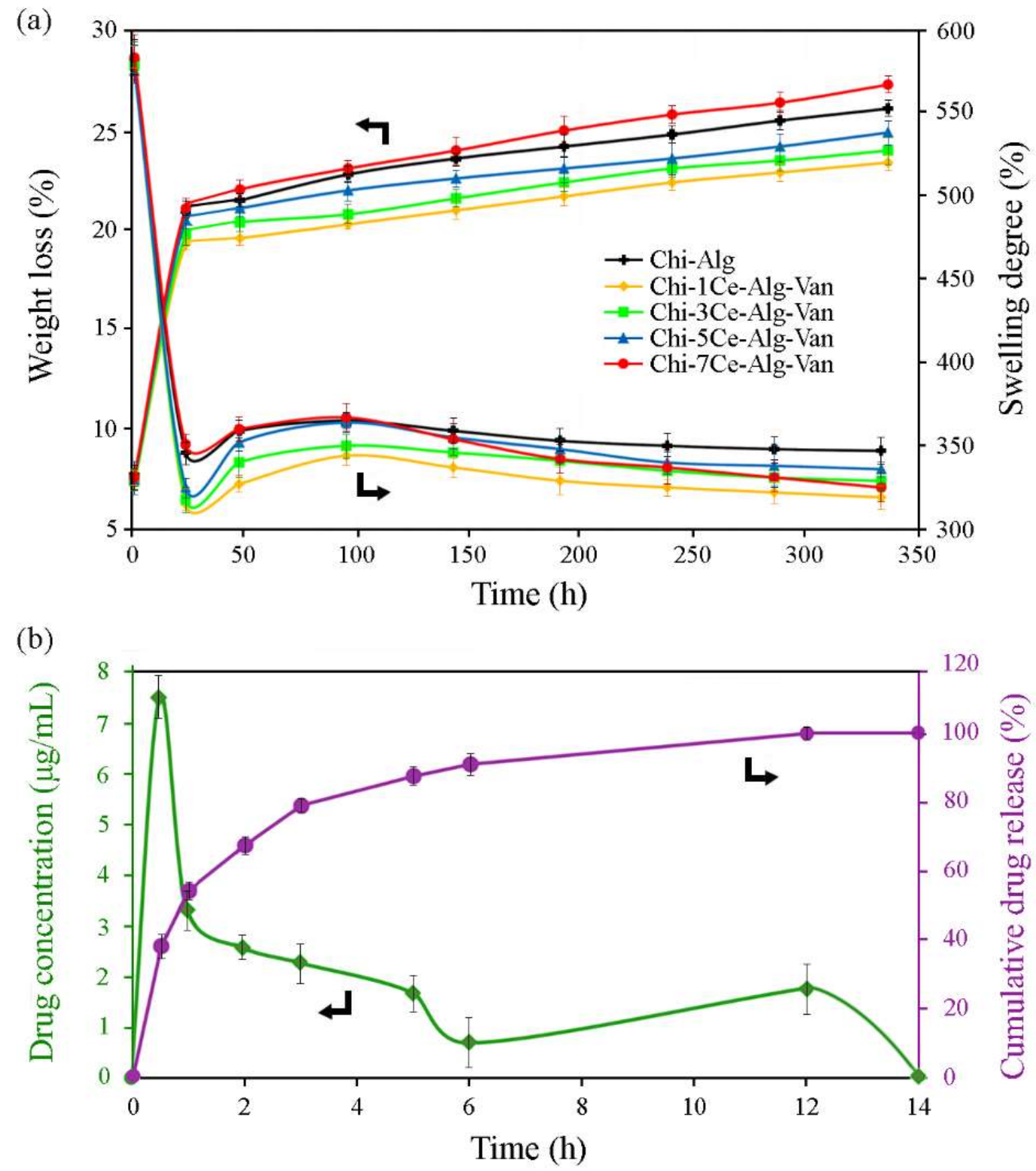

*Fig. 3. Degradation and swelling behaviors of the different scaffolds (a) and vancomycin release kinetics (b) of Chi-5Ce-Alg-Van.*

At initial hours of immersion in PBS, the weight of all the scaffolds rapidly decreases and reaches a remarkable weight loss of around 8% by only 3 h. This is attributed to the initial rapid degradation of alginate, which is water-soluble. This level of degradability is promising for the initial burst release of vancomycin into the wound bed as it can create a shock effect and eliminate any pre-existing bacteria [43]. Furthermore, the rapid degradation of alginate beneficially exposes chitosan to the medium, providing more effective hemostasis and cellular activities [44]. After one day of immersion, the rate of weight loss gradually decreases for all the samples, and after 3 days, it stabilizes at a relatively constant value of approximately 1% per day. These rates shows potential for the sustained and long-term release of the cerium oxide nanoparticles, which is necessary to provide effective antioxidant properties in the wound bed until healing is achieved [7]. The gradual degradation of the scaffolds also allows them to maintain their structural integrity and to support the wound for a longer period, ultimately facilitating wound healing and the formation of new tissue [45]. Typically, the achieved weight loss of approximately 25% after 14 days in this study aligns with degradation results reported in some successful studies aimed at treating chronic wounds [28,46].

The wet weight of the scaffolds gradually increases over time, giving a maximum swelling degree of approximately 580 % after 1 h of immersion in PBS. This indicates that the scaffolds have absorbed water approximately 6.7 times their original dry weight. The degree of swelling decreases gradually with further immersion in PBS and ultimately reaches equilibrium after 7 days. The equilibrium swelling level of the scaffolds is approximately 330%, indicative of their capability to absorb a significant amount of fluid. This is essential for maintaining a moist environment conducive to wound healing [47], removing wound exudates [48], and preventing bacterial accumulation [49] .The considerable swelling capacity of the scaffolds is related to the hydrophilic nature of both alginate and chitosan. Alginate is a water-soluble polysaccharide that can absorb large amounts of water due to its ability to form hydrogen bonds with water molecules [50]. Chitosan is also a hydrophilic polymer that can interact with water molecules through electrostatic interactions [51]. The initial rapid swelling behavior of the scaffolds is attributed to the presence of the alginate coating on their surface, which can absorb and retain a significant amount of fluid in the initial hours. After the degradation of the alginate coating in the initial hours, a reduction in the swelling of the scaffolds is observed. The swelling capability of the scaffolds in this step is associated with chitosan. Overall, it can be concluded that the fabricated scaffolds exhibit a significant swelling behavior and are suitable for wound dressing applications from this perspective. Wound dressings with superior swelling properties can absorb wound exudate, thereby reducing bacterial infection and aiding in wound healing. The swelling degree of approximately 330% obtained in this work aligns with the results of successful studies, indicating that swelling degrees ranging within 200-400% are suitable for wound dressing applications [8,52].

### *3.4. Vancomycin release kinetics of dressings*

The release behavior of vancomycin from the dressings is depicted in Fig. 3b, exhibiting a burst release of almost 60% of loaded vancomycin within the first hour. This release mode is in agreement with the significant swelling of the dressing due to the presence of the alginate hydrogel coating as the first release mechanism. Also, approximately 91% of vancomycin is released in 6 h, which aligns with the rapid degradation of the alginate coating during the initial hours as the second release mechanism of vancomycin loaded. These release mechanism assignments are is in agreement with those described in Ref. [27] focusing on the two-stage release of silver sulfadiazine from alginate coatings deposited on chitosan nanogels.

The initial burst release behavior of vancomycin is promising for treating chronic wounds as it can create a shock effect on bacteria, effectively eliminating them and promoting subsequent healing processes [43]. Some infections start by the formation of a protective biofilm by pathogenic bacteria within 10 h, shielding them from antibiotics and immune cells. Therefore, in early stages of wound treatment, the rapid and localized release of antibiotics is crucial to prevent these complications [53]. Regarding concentration, it has been shown that a vancomycin concentration of 5.0-10.0 µg/mL within the first 24 h is optimal to provide strong antibacterial effects without causing significant cytotoxicity [52,54]. This suggests the appropriate concentration of vancomycin released from the dressings developed in this study, characterized by 7.5 µg/mL within 24 h.

### *3.5. Antibacterial activity of dressings*

As displayed in Fig. 4a-b, the diameter of the growth inhibition zone against *S. aureus* and *E. coli* for all the tested samples are consistently around 26 and 0 mm, respectively. Previous studies on antibacterial dressings for chronic wounds management indicate that inhibition zones ranging within 20-29 mm correspond to robust antibacterial activity [22,55–57]. Accordingly, the antibacterial efficacy of the scaffold dressings developed in this work is considered suitable for treating chronic wounds infected by *S. aureus*. The burst release of vancomycin in the early hours, as shown in Fig. 3b, facilitates the effective elimination of gram-positive bacteria. Typically, vancomycin, a glycopeptide antibiotic, interferes with bacterial cell wall synthesis by binding to the D-alanyl-D-alanine portion of peptidoglycan precursors, preventing their incorporation into the cell wall and thereby weakening the bacterial structure [58,59]. It is noteworthy that vancomycin is only effective against gram-positive bacteria. The outer membrane structure of gram-negative bacteria, which consists of two lipid layers, acts as an additional barrier that prevents vancomycin from penetrating and exerting its antibacterial effects [3]. As a result, the scaffolds were ineffective against *E. coli* bacteria (Fig. 4b). It is also worth mentioning that *S. aureus* gram-positive bacterium is the primary cause of severe skin infections and biofilm formation [60,61]. As a result, the antibacterial effects of the produced wound dressings against this bacterium potentially suffices for wound dressing applications.

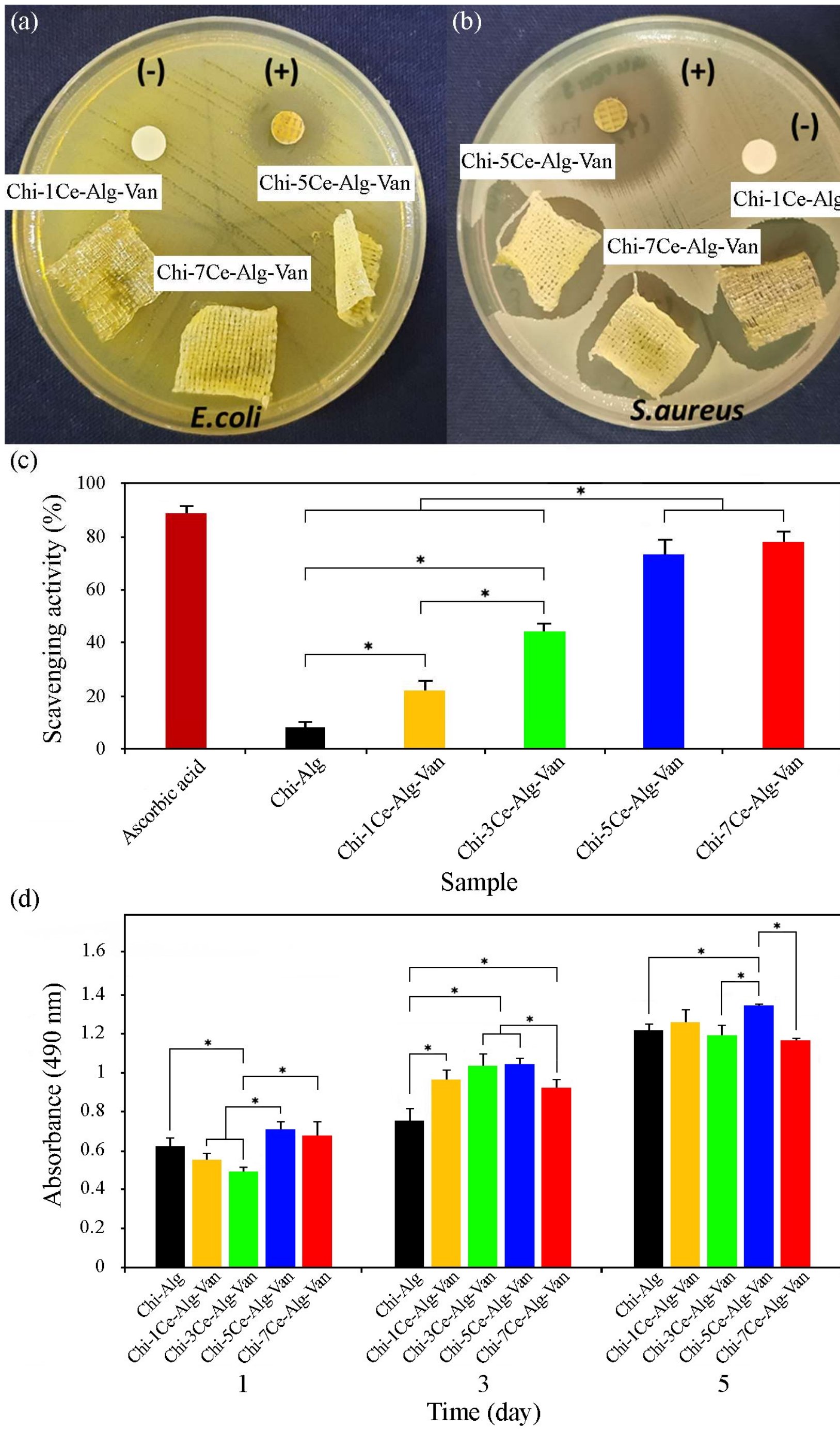

*Fig. 4. Antibacterial disk diffusion assay results of the samples against E. coli (a) and S. aureus (b), DPPH free radical scavenging activity of the samples at a concentration of 25 mg/mL (c), and MTS assay results (d), with * denoting statistically significant differences with p<0.05.*

The absence of significant differences (p-values < 0.05) in the inhibition zone diameter of the samples containing the various concentrations of cerium oxide denies the antibacterial contribution of the cerium oxide nanoparticles. This contrasts with observations in some studies regarding the antibacterial activity of cerium oxide nanoparticles against *S. aureus* [62–64]. These differences arise from the diversity in cerium oxide concentrations used. Considerable antibacterial effects of cerium oxide are observed at concentrations above 200 µg/mL, which also comes with cytotoxic effects [65]. However, in this study, lower concentrations suitable were investigated for antioxidant properties and enhanced cellular viability.

### *3.6. Antioxidant capability of dressings*

The DPPH radical inhibition percentage of the samples is indicated in Fig. 4c. Chi-Alg shows an antioxidant activity of almost 7.8%, due to the weak antioxidant property of chitosan. But the antioxidant property significantly increases with the addition of the cerium oxide nanoparticles. Typically, Chi-1Ce-Alg-Van with only one percent of cerium oxide exhibits almost 21.9% scavenging activity with a statistically significant difference (p<0.05) with respect to Chi-Alg. This trendline continues as Chi-3Ce-Alg-Van, Chi-5Ce-Alg-Van, and Chi-7Ce-Alg-Van present almost 44.2, 73.4, and 78.1% scavenging ability, respectively. However, the statistical analysis of the data revealed that there is no significant difference between Chi-5Ce-Alg-Van and Chi-7Ce-Alg-Van.

The production of ROS inside cells by oxygen-generating species is essential for various biochemical processes, but excessive production can disrupt these processes. In the inflammatory stage, ROS plays a crucial role in wound healing, but an increase in free radicals at the wound site leads to oxidative tissue damage and disruption of wound healing [18]. Antioxidant substances regulate the amount of free radicals, preventing oxidative stress and cellular tissue damage as well as increasing cell survival and proliferation. Cerium oxide nanoparticles, by reversibly transitioning between reduction ($Ce^{3+}$) and oxidation ($Ce^{4+}$) states, can remove ROS and cellular antioxidant stress [66]. Also, cerium oxide nanoparticles increase the expression of superoxide dismutase 1, an enzyme that reduces oxidative stress at low concentrations [17].

Zeng *et al.* [7] reported a DPPH scavenging ability of approximately 60% by antioxidant and antibacterial wound dressings with antioxidant epigallocatechin gallate. Also, Augustine *et al.* [67] reported the DPPH scavenging effectiveness of 45% for gelatin methacryloyl dressings containing 4 wt% of cerium oxide nanoparticles. Furthermore, Ye *et al.* [68] pointed out chitosan-polyethylene glycol hydrogels containing epigallocatechin gallate and cerium oxide complexes, demonstrating a DPPH scavenging ability of around 50%. That is, in comparison to the literature, the Chi-5Ce-Alg-Van and Chi-7Ce-Alg-Van dressings developed in this research possess stronger antioxidant properties. This predominance can be due to the biphasic release kinetics of the cerium oxide nanoparticles, resulting from the initial swelling of chitosan followed by its slow degradation [69]. Indeed, due to the high level of ROS in early stages, a greater release of cerium oxide is required. Subsequently, a sustained and long-term release is necessary to eliminate excess ROS and prevent cellular cytotoxicity until the wound heals [19,70].

### *3.7. Cell cytocompatibility of dressings*

The MTS cell metabolic activity assay results are reparented in Fig. 4d, indicating considerable cellular proliferation on all the samples with culture time, due to the favorable biological activities of the dressings. From a comparative perspective, the cell metabolic activity increases with the concentration of cerium oxide up to 5% throughout the entire culture period, but dropped when the level reaches 7%. The enhancement is due to the positive effect of ROS inhibition by cerium oxide on cell metabolic activity, where ROS accumulation causes oxidative damage and negatively affects cellular proliferation [71]. The following reduction in the cell metabolic activity, particularly on days 3 and 5, can be attributed to the extreme release of cerium oxide. At high concentrations, cerium oxide induces cytotoxicity by disrupting cellular antioxidant systems, resulting in increased ROS levels. Additionally, the excessive expression of pro-inflammatory factors further contributes to the cellular toxicity caused by cerium oxide [72]. In the first day, the slight reduction in the cell metabolic activity of the wound dressings containing vancomycin compared to the control can be related to the high initial burst release of vancomycin [73]. However, this negative effect is not observed in the dressings containing 5% and 7% cerium oxide, due to the positive effects of cerium oxide on cell metabolic activity.

Zeng *et al.* [7] pointed out that hydrogel wound dressings containing epigallocatechin gallate and copper, with antioxidant and antibacterial properties, showed a cell metabolic activity of 106% after 5 days. In another study, Bagheri *et al.* [74] reported chitosan-based dressings containing silver and zinc oxide with antibacterial and antioxidant properties presented a cell metabolic activity of approximately 100% by 5 days. In several studies focused on developing dressings with either antibacterial or antioxidant properties, the cell metabolic activity ranging from 100% to 112% after 5 days has been reported [75,76]. Therefore, it is concluded that the optimal dressing developed in this research ranks highly among similar dressings due to its superior antibacterial and antioxidant properties, as well as excellent cell cytocompatibility, with a cell metabolic activity of 109.8% after 5 days with respect to the antibiotic- and antioxidant-free sample (Chi-Alg).

### *3.8. Cell migration promotion of dressings*

Fig. 5 illustrates the scratch assay results of the dressings, indicating that the higher cerium oxide content corresponds to increased cell migration rates. Remarkably, after 24 h, both the Chi-5Ce-Alg-Van and Chi-7Ce-Alg-Van samples completely fill the scratched area, suggesting the high potential of these dressings in facilitating wound closure [77]. According to the literature [19,74,78], wound closure percentages above 80% after 24 hours are considered satisfactory, indicating that the wound dressings developed in this research achieve the ideal result of 100% wound closure in this period.

During wound healing, cell migration, particularly by fibroblasts, is crucial. Fibroblasts migrate into the wound area in response to signals that trigger them to secrete extracellular matrix (ECM) proteins and proliferate for tissue regeneration and wound closure [79]. As the wound healing process progresses, cells surrounding the wound collectively migrate across the provisional matrix rich in protein and growth factors during re-epithelialization. The leading cells adjust their expression of adhesion proteins to facilitate migration through this matrix, contributing to the closure of the wound [80]. In these processes, the presence of cerium oxide reduces oxidative stress at the wound site and potentially accelerates wound healing by increasing cell migration and proliferation [17]. The role of chitosan and alginate is also critical in terms of inducing the proliferation of dermal fibroblasts, stimulating fibroblast cell migration, and collagen deposition [81]. Chitosan also positively influences cell migration because of its N-acetyl-D-glucosamine units, which are the primary glycosaminoglycans found in ECM [82].

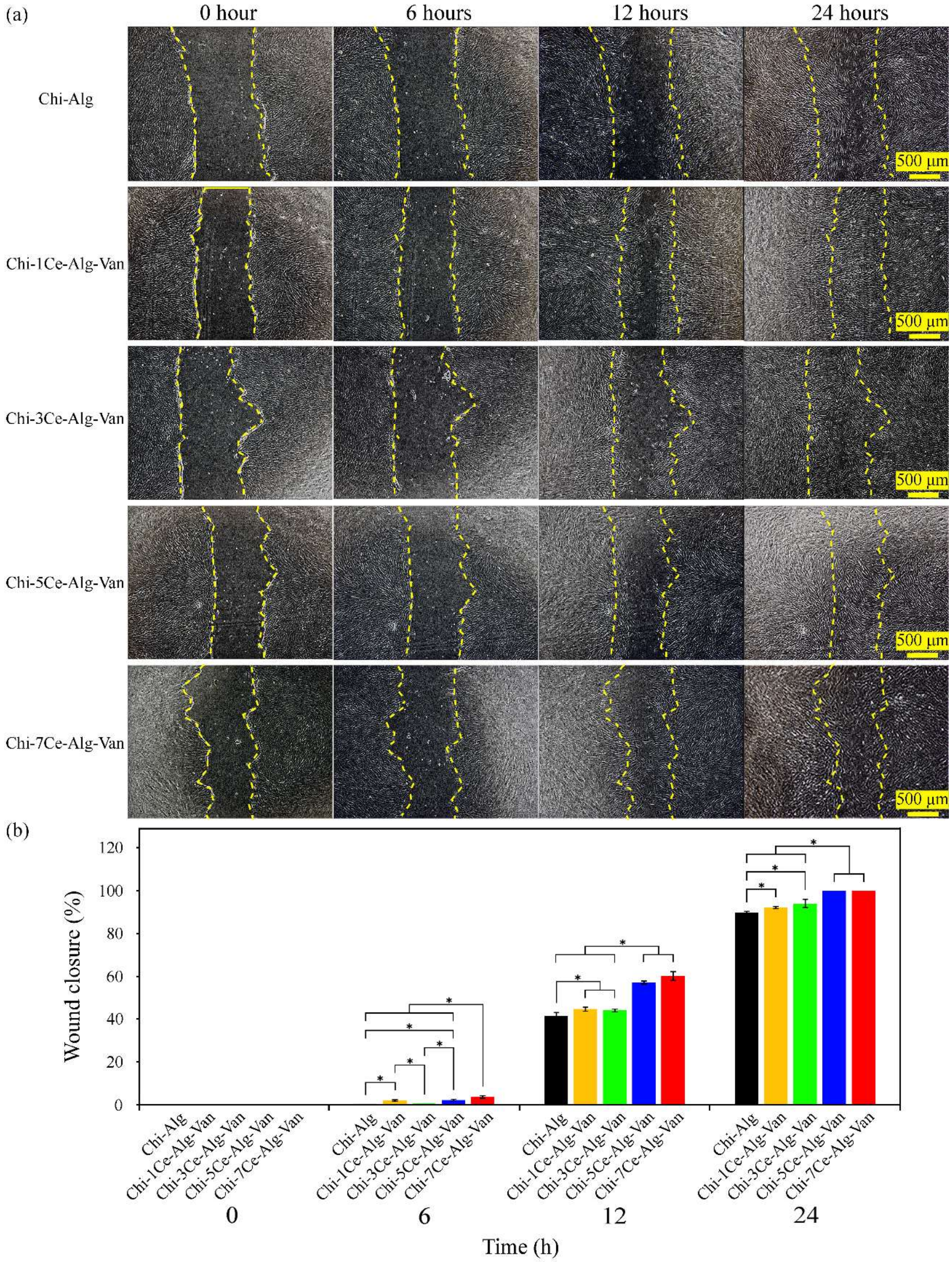

*Fig. 5. Scratch assay results of the dressings: representative photomicrographs (a) and percentage of wound surface closure, with * indicating significant differences (*p<0.05) (b).*

## 4. Conclusion

In this work, chitosan-cerium oxide dressings with an alginate-vancomycin coating layer were developed for the treatment of chronic wounds. The incorporation of cerium oxide nanoparticles and vancomycin into the dressings improved their antioxidant, antibacterial, cell metabolic, and cell migration characteristics, as tabulated in Table 2. Approximately 100% of loaded vancomycin was released within 12 h, aligning with the rapid degradation and high swelling rate of alginate during the initial hours. Due to this burst release of vancomycin, the dressings exhibited strong antibacterial properties against *S. aureus* bacteria, as evidenced by large growth inhibition zones of 26 mm observed in the disk diffusion assay. The cerium oxide containing dressings also demonstrated a strong antioxidant capacity, with the antioxidant activity increasing with the concentrations of the cerium oxide nanoparticles in the dressings. Typically, the Chi-5Ce-Alg-Van formulation possessed the highest cell cytocompatibility with a cell metabolic activity of 109.8% by 5 days, over 73.4% DPPH scavenging ability, and 100% wound closure potential within 24 h. Conclusively, the optimal dressings developed in this research provide an enhanced balance of antibacterial and antioxidant properties, along with excellent cell activity characteristics, making them a promising candidate for chronic wounds management.

*Table 2. Comparative properties of the different dressings*

| Sample | Antibacterial inhibition zone (mm) | Scavenging activity (%) | Cell viability (%) | Wound closure (%) |
|---|---|---|---|---|

| | **Species** | | - | **Time (day)** | | | **Time (h)** | | |
|---|---|---|---|---|---|---|---|---|---|
| | ***S. aureus*** | ***E. coli*** | | **1** | **3** | **5** | **6** | **12** | **24** |
| **Chi-Alg** | - | - | 7.8 | 100.0 | 100.0 | 100.0 | 0.1 | 41.3 | 89.6 |
| **Chi-1Ce-Alg-Van** | 26 | 0 | 21.9 | 88.0 | 129.3 | 103.3 | 1.6 | 44.5 | 92.0 |
| **Chi--3Ce-Alg-Van** | - | - | 44.2 | 78.4 | 138.7 | 97.6 | 0.3 | 43.9 | 93.9 |
| **Chi-5Ce-Alg-Van** | 26 | 0 | 73.4 | 113.6 | 140.0 | 109.8 | 1.9 | 57.0 | 100.0 |
| **Chi-7Ce-Alg-Van** | 26 | 0 | 78.1 | 108.8 | 123.9 | 95.8 | 3.4 | 60.0 | 100.0 |

**Acknowledgment**

L.T. acknowledges the support from National Institutes of Health under award numbers R56 DE029191 and 1R21EY035480-01.